# Ringworlds and Dyson spheres can be stable

Colin R. McInnes★
*James Watt School of Engineering, University of Glasgow, Glasgow G12 8QQ, UK*

**ABSTRACT**
In his 1856 Adams Prize essay, James Clark Maxwell demonstrated that Saturn's rings cannot be comprised of a uniform rigid body. This is a consequence of the two-body gravitational interaction between a ring and planet resulting in instability. Similarly, it is also known that a so-called Dyson sphere encompassing a single star would be unstable due to Newton's shell theorem. A surprising finding is reported here that both a ring and a sphere (shell) can be stable in the restricted three-body problem. First, if two primary masses are considered in orbit about their common centre of mass, a large, uniform, infinitesimal ring enclosing the smaller of the masses can in principle be stable under certain conditions. Similarly, a Dyson sphere can, be stable, if the sphere encloses the smaller of the two primary masses, again under certain conditions. These findings extend Maxwell's results on the dynamics of rings and have an interesting bearing on so-called Ringworlds and Dyson spheres from fiction. Moreover, the existence of passively stable orbits for such large-scale structures may have implications for so-called techno-signatures in search for extra-terrestrial intelligence studies.

## 1. INTRODUCTION

The circular restricted three-body problem has five well-known equilibrium solutions comprising three unstable collinear equilibria and two triangular equilibria, which may be stable if the mass ratio of the problem is less than Routh's value (Battin, 1999). There exists a range of extensions to the circular restricted three-body problem to more complex cases. These include the effect of radiation pressure from one or both primary masses (Simmons, McDonald and Brown, 1985), one or more oblate primary masses (Sharma and Rao, 1975) and the consideration of a non-point infinitesimal mass (Singh and Leke, 2014). Other models include the addition of a fixed annular ring encompassing one of the primary masses (Xiangling, 1982) or encompassing both of the primary masses to model a circumstellar dust belt (Chakraborty, Narayan and Ishwar, 2021). A fluid-filled sphere has been considered to investigate the interaction of a planetary moon and planetary core (Robe, 1977), while the more general problem of two primary masses and a homogenous sphere has also been investigated (Bozis and Michalodimitrakis, 1982).

These extensions to the classical restricted three-body problem follow various extensions to the restricted two-body problem comprising an infinitesimal body in orbit about a single primary mass. Extensions include a non-point infinitesimal mass, such as a uniform solid ring enclosing the primary mass. Maxwell (1859) demonstrated that a single equilibrium configuration exists, where the centre of the solid ring is superimposed on the primary mass, but that the equilibrium configuration is unstable (whilst also acknowledging the findings of Laplace on ring instability (Whiting, 2011)). Maxwell also noted that the equilibrium configuration can be linearly stable if a massive particle is attached to the ring. Further analysis on the perturbed ring problem is provided by Pendse (1935). Apparently Herschel also speculated that a non-uniform ring could potentially be stable, while Laplace also investigated the dynamics of the problem (Whiting, 2011).

The stability of an artificial flexible ring enclosing a central mass was investigated by Breakwell (1981). McInnes investigated the non-linear ring problem by determining the gravitational force on a solid ring at an arbitrary displacement from the central mass (McInnes, 2003). It was shown that the sole equilibrium solution was unstable as expected. Furthermore, it was noted that while the equilibrium configuration is unstable for in-plane displacements, in principle out-of-plane displacements are stable for a range of initial conditions. Such out-of-plane displacements were discussed in detail by Schumayer and Hutchinson (2019) and indeed noted by Maxwell (1859). Rippet (2014) considered the coupled orbit and attitude dynamics of a massive rotating and precessing ring. Others have also considered the orbits of an infinitesimal point mass in the vicinity of a massive ring. For example Alberti and Vidal (2007) consider a homogeneous annular disk, with a range of symmetries noted.

In this paper, the restricted three-body problem will be considered with a uniform ring of infinitesimal mass to form a ring-restricted three-body problem. It will be assumed that the ring orientation is fixed in a rotating frame of reference with the axis of symmetry of the ring normal to the orbit plane. There are then three possible configurations for a ring in the restricted three-body problem: with the ring enclosing both primary masses, one primary mass, or none. The ring-restricted three-body problem is quite distinct from the classical restricted three-body problem due to the topology of the ring. Since the centre of mass of the ring is empty, it can in principle be superimposed over one of the primary masses without resulting in a singularity. A singularity will occur, however, if the ring itself contacts a primary mass. This leads to an extended collision set that can be mapped on to the geometry of the problem. There are similarities with the relative equilibria of the three-body problem with finite-sized masses, where the masses can be in contact without leading to a singularity (Scheeres, 2016).

★ Email: colin.mcinnes@glasgow.ac.uk

It will be shown that up to seven equilibria can now exist, with five collinear equilibria and two triangular equilibria. The two additional collinear equilibria are new to the ring-restricted three-body problem. The linear stability properties of the equilibria will be explored as a function of the mass ratio of the problem and the non-dimensional radius of the ring. It will be shown that Routh's criteria for stability of the triangular equilibrium points can be extended to include the non-dimensional ring radius. Moreover, unlike the classical restricted three-body problem, under certain conditions, collinear equilibria can be stable for a (small) range of mass ratios and a range of ring radii. In particular, it will be shown that a uniform solid ring can be in a linearly stable equilibrium configuration while enclosing the smaller of the primary masses. This extends the findings of Maxwell on the two-body stability of a uniform ring, demonstrating that rings can, in principle, be stable in a restricted three-body problem.

There is a range of natural ring-like structures, such as self-gravitating dust rings (Sparkle, 1986), ring-like galaxies (Bannikova, 2018), and accretion discs in binary star systems. However, more speculative applications of the ring-restricted three-body include orbital loops (Polyakov, 1977; Birch, 1982), space habitats (Johnson and Holbrow, 1977), and ringworlds (Niven, 1970, 1980; Raval and Srikanth 2024) and orbitals (Banks, 1987) from fiction. Moreover, due to contemporary interest in techno-signatures in SETI studies, the analysis presented provides insights into the dynamics of ultra-large, artificial space structures that could support future observational searches (Wright et al. 2022). The material limits of such structures are not considered, although such considerations are clearly important (Fridman et al. 1984).

In addition to the ring-restricted three-body problem, its extension to a shell-restricted three-body problem will also be investigated. Here, the (1D) closed ring is replaced by a (2D) closed shell, again of infinitesimal mass. The problem can be investigated using Newton's shell theorem (Reed, 2022) with three possible configurations: with the shell enclosing both primary masses, one primary mass, or none. Similar to the ring-restricted three-body problem, it will be demonstrated that two new collinear equilibria exist and can be linearly stable under certain conditions. The problem extends the concept of a Dyson sphere encompassing a single star (Dyson, 1960) to binary star systems. Again, the analysis presented here could help support observational searches for such large-scale, artificial space structures (Suazo et al. 2022, 2024; Wright et al. 2022), although material limits are not considered. A related investigation of natural shell-like structures has also considered the restricted three-body problem embedded within an extended halo of matter (Mia et al. 2024).

The paper is organized as follows. Section 2 defines the ring-restricted three-body problem with two primary masses and an infinitesimal solid ring. The conditions for equilibria in the ring-restricted three-body problem are determined in Section 3, with two new equilibria identified, while the Jacobi integral of the problem is determined in Section 4. The linear stability properties of the equilibria are investigated in Section 5, with an extension of Routh's criteria found for the triangular points and the conditions for stable collinear equilibria investigated. An extension of the ring-restricted three-body problem to hollow shells is presented in Section 6. Finally, a range of applications of the ring and shell-restricted three-body problems are discussed in Section 7, while conclusions are drawn in Section 8.

## 2. RING-RESTRICTED THREE-BODY PROBLEM

A rotating frame of reference will now be considered with angular velocity $\boldsymbol{\omega}$ directed along the $\boldsymbol{e}_3$ axis, as shown in Fig. 1. Then, two primary masses $m_1$ and $m_2$ will be located along the $\boldsymbol{e}_1$ axis of the rotating frame with a uniform solid ring of mass $m$ located within the $\boldsymbol{e}_1 - \boldsymbol{e}_2$ plane. As will be seen, equilibria are found in the $\boldsymbol{e}_1 - \boldsymbol{e}_2$ plane, although infinitesimal out-of-plane displacements will be considered for the linear stability analysis. It will be assumed that the ring orientation is invariant and fixed in a rotating frame of reference with the axis of symmetry of the ring normal to the orbit plane. Moreover, rotation about the ring normal will not effect its orbital motion, due to the ring's symmetry.

For a restricted three-body problem it is assumed that $m << m_1, m_2$. In the usual manner for the restricted three-body problem (Battin, 1999) the angular velocity of the rotating frame of reference will be defined by the orbital angular velocity of the two primary masses about their common centre-of-mass. Moreover, a non-dimensional mass ratio will be defined such that $\mu = m_2/m_1 + m_2$, with $m_1 + m_2 = 1$, so that $0 < \mu \leq 1/2$. The unit of length will be defined by the separation of the primary masses, while the unit of time will be chosen such that $|\boldsymbol{\omega}| = 1$. The two primary masses are then located at positions $\bar{\boldsymbol{r}}_1$ and $\bar{\boldsymbol{r}}_2$ given by $(-\mu, 0, 0)$ and $(1 - \mu, 0, 0)$, respectively, again shown in Fig. 1.

The gravitational potential $\Phi(\bar{\boldsymbol{r}}, \boldsymbol{r})$ between a uniform ring with centre $C$ at position $\boldsymbol{r}$ and a point mass at position $\bar{\boldsymbol{r}}$ is determined in Appendix A. The potential due to the two primary masses $m_1$ and $m_2$ will then be defined as $\Phi_1(\bar{\boldsymbol{r}}_1, \boldsymbol{r})$ and $\Phi_2(\bar{\boldsymbol{r}}_2, \boldsymbol{r})$, respectively, and the total potential energy is then given by $\Phi = \Phi_1(\bar{\boldsymbol{r}}_1, \boldsymbol{r}) + \Phi_2(\bar{\boldsymbol{r}}_2, \boldsymbol{r})$. Moreover, the centripetal and Coriolis accelerations acting on the ring in the rotating frame of reference are also determined in Appendix A. The equation of motion of the ring in the rotating frame of reference can therefore be written as

$$\ddot{\boldsymbol{r}} + 2\boldsymbol{\omega} \times \dot{\boldsymbol{r}} + \boldsymbol{\omega} \times (\boldsymbol{\omega} \times \boldsymbol{r}) = -\nabla \Phi, \quad (1)$$

where the acceleration due to the primary masses $m_1$ and $m_2$ can be written (from Appendix A) as

$$\nabla \Phi_1(\bar{\boldsymbol{r}}_1, \boldsymbol{r}) = -\frac{2(1-\mu)}{\pi R} \frac{\partial}{\partial \boldsymbol{r}} \left[ \frac{1}{k\sqrt{1-\gamma}} K(\Lambda) \right]_{\bar{\boldsymbol{r}}=\bar{\boldsymbol{r}}_1} \quad (2a)$$

$$\nabla \Phi_2(\bar{\boldsymbol{r}}_2, \boldsymbol{r}) = -\frac{2\mu}{\pi R} \frac{\partial}{\partial \boldsymbol{r}} \left[ \frac{1}{k\sqrt{1-\gamma}} K(\Lambda) \right]_{\bar{\boldsymbol{r}}=\bar{\boldsymbol{r}}_2}. \quad (2b)$$

where $K(\Lambda)$ is a complete elliptical integral of the first kind, $k = \sqrt{1 + (\boldsymbol{r} - \bar{\boldsymbol{r}})^2/R^2}$, $\gamma = \sqrt{k_1^2 + k_2^2}$ where $k_1 = 2(\boldsymbol{r} - \bar{\boldsymbol{r}}) \cdot \boldsymbol{e}_1/Rk^2$ and $k_2 = 2(\boldsymbol{r} - \bar{\boldsymbol{r}}) \cdot \boldsymbol{e}_2/Rk^2$, $\Lambda = 2\gamma/(\gamma - 1)$ and $R$ is now the non-dimensional ring radius, as detailed in Appendix A.

In order to proceed, the centripetal acceleration in equation (1) can be written as the gradient of a potential $U$ such that:

$$\nabla U = \boldsymbol{\omega} \times (\boldsymbol{\omega} \times \boldsymbol{r}), \quad U = -\frac{1}{2}|\boldsymbol{\omega} \times \boldsymbol{r}|^2 \quad (3)$$

and so an augment potential can now be defined as $V = U + \Phi$. Then, equation (3) can be written more compactly as

$$\ddot{\boldsymbol{r}} + 2\boldsymbol{\omega} \times \dot{\boldsymbol{r}} + \nabla V = 0. \quad (4)$$

Before investigating the location of equilibria in the ring-restricted three-body problem, collision sets $\Sigma_1$ and $\Sigma_2$ can be defined where

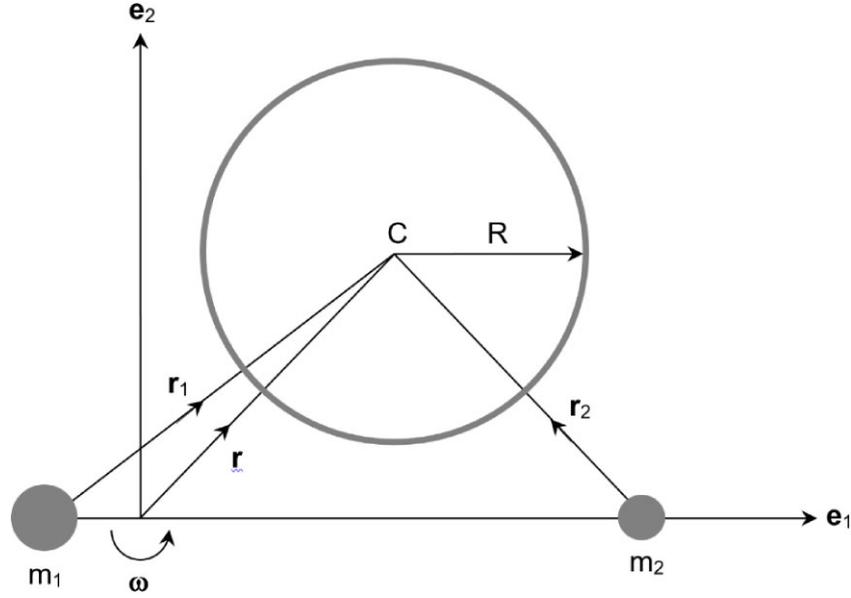

**Figure 1.** Ring-restricted three-body problem with primary masses $m_1$ and $m_2$ in a frame of reference rotating with angular velocity $\omega$ and a uniform ring of radius $R$ with centre.

the ring intersects either of the primary masses $m_1$ or $m_2$ such that for in-plane motion:

$$\Sigma_1 = \left(r \in \mathbb{R}^2 : (x + \mu)^2 + y^2 = R^2\right) \tag{5a}$$

$$\Sigma_2 = \left(r \in \mathbb{R}^2 : (x - (1 - \mu))^2 + y^2 = R^2\right). \tag{5b}$$

The planar ring-restricted three-body problem is therefore defined on $\mathbb{R}^2 - (\Sigma_1 \cup \Sigma_2)$. Since the mass of the ring-restricted three-body problem is extended, rather than a point mass, the singularities of the problem are represented by rings. The centre-of-mass of the ring is empty, and so can enclose one (or both) of the primary masses. Therefore, there are no singularities when the centre of the ring is superimposed on a primary masses (unless the radius of the ring radius is equal to the separation of the primary masses). Now that the ring-restricted three-body problem has been defined, along with its collision sets, the conditions for equilibria will be investigated.

## 3. CONDITIONS FOR EQUILIBRIUM

The dynamics of the ring-restricted three-body problem defined by equation (4) can now be used to determine the conditions for equilibrium configurations of the ring relative to the two primary masses in the rotating frame of reference. From equation (4) the condition for an equilibrium configuration with $\ddot{r} = \dot{r} = 0$ is given simply by $\nabla V = 0$ which can then be written as

$$\omega \times (\omega \times r) - \frac{(1-\mu)}{\pi R} \frac{\partial}{\partial r}\left[\frac{1}{k_1\sqrt{1-\gamma}} K(\Lambda)\right]_{\bar{r}=\bar{r}_1}$$
$$- \frac{\mu}{\pi R} \frac{\partial}{\partial r}\left[\frac{1}{k_1\sqrt{1-\gamma}} K(\Lambda)\right]_{\bar{r}=\bar{r}_2} = 0, \tag{6}$$

where again $\bar{r}_1 = (-\mu, 0, 0)$ and $\bar{r}_2 = (1 - \mu, 0, 0)$. It can be demonstrated that $\nabla V \cdot e_2 = 0$ when $y = 0$ and $\nabla V \cdot e_3 = 0$ when $z = 0$, as expected due to symmetry since $V(-y) = V(y)$ and $V(-z) = V(z)$. Therefore, since the condition for an equilibrium configuration is given by $\nabla V = 0$; it can be expected that all equilibria will lie in the $e_1 - e_2$ plane and a set of collinear equilibria can be expected to exist along the $e_1$ axis. Other equilibria with $\nabla V \cdot e_1 = 0$ and $\nabla V \cdot e_2 = 0$ can be expected to exist for $y \neq 0$. In searching for equilibria using by equation (6) the search space can be reduced using the above symmetries.

Given the complexity of the potential of the problem analytical solutions for the location of equilibria do not appear to be available. It is noted, however, that approximate solutions may be available for small ring radii and/or small mass ratios, although this is not considered further here. Numerically, it is found from equation (6) that the ring-restricted three-body problem possesses seven equilibrium solutions, five collinear points, and two triangular points. The collinear points are to be expected since $\nabla V \cdot e_2 = 0$ when $y = 0$ and $\nabla V \cdot e_3 = 0$ when $z = 0$. These solutions comprise three equilibria analogous to the $L_1$, $L_2$, and $L_3$ points of the classical restricted three-body problem. However, there are two addition collinear points, defined here as $L_6$ and $L_7$ which in general are located close to the two primary masses $m_1$ and $m_2$. Similarly, the triangular points are analogous to the $L_4$ and $L_5$ points when $\nabla V \cdot e_1 = 0$ and $\nabla V \cdot e_2 = 0$ for $y \neq 0$.

For illustration, the location of the seven equilibria and the collision sets $\Sigma_1$ and $\Sigma_2$ are shown in Fig. 2 for the equal mass ring-restricted three-body problem with $\mu = 1/2$ and with unit ring radius $R = 1$. The location and properties of the equilibria are listed in Table 1. Again, the collisions sets correspond to the set of points where, if the ring centre were placed there, the ring would contact a primary mass resulting in a singularity. For example, the centre of the dashed ring on the left of Fig. 2 is located on $\Sigma_1$ and so the ring intersects $m_1$, while the dashed ring on the right of Fig. 2 is not located on $\Sigma_2$ and so the ring does not intersect $m_2$.

In order to illustrate the configuration of the ring and primary masses, the ring is superimposed on the seven equilibria in Fig. 3. It can be seen from Fig. 3a that at the $L_1$ point the ring encloses both primary masses, while at the new $L_6$ and $L_7$ points the ring encloses one of the primary masses. From Fig. 3b, it can seen that at the $L_2$, $L_3$, $L_4$, and $L_5$ points the ring does not enclose either of the primary masses. Therefore, the collision sets will only constrain the motion of the ring in the vicinity of $L_1$, $L_6$, and $L_7$.

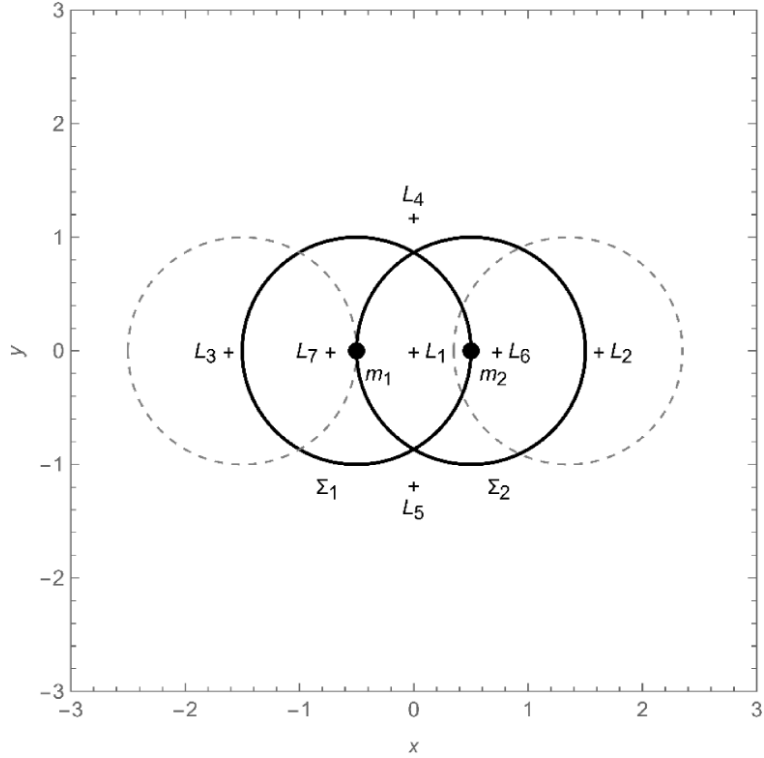

**Figure 2.** Location of the seven equilibrium points for the equal mass problem with $\mu = 1/2$ and unit ring radius $R = 1$. The collision sets are defined by $\Sigma_1$ and $\Sigma_2$.

**Table 1.** Location and associated properties of the equilibrium configurations for the equal mass problem with $\mu = 1/2$ and unit ring radius $R = 1$.

| Point | x | y | z | Eigenvalues | V |
|---|---|---|---|---|---|
| $L_1$ | 0 | 0 | 0 | $\pm(0.96796 + 0.99577i)$<br>$\pm(0.96796 - 0.99577i)$<br>$\pm 1.37504i$ | $-1.07318$ |
| $L_2$ | $+1.61936$ | 0 | 0 | $\pm 3.04113,$<br>$\pm 0.74271i$<br>$\pm 3.27060i$ | $-2.20282$ |
| $L_3$ | $-1.61936$ | 0 | 0 | $\pm 3.04113$<br>$\pm 0.74271i$<br>$\pm 3.27060i$ | $-2.20282$ |
| $L_4$ | 0 | $+1.18008$ | 0 | $\pm(1.18529 + 0.77842i)$<br>$\pm(1.18529 - 0.77842i)$<br>$\pm 1.89683i$ | $-1.66947$ |
| $L_5$ | 0 | $-1.18008$ | 0 | $\pm(1.18529 + 0.77842i)$<br>$\pm(1.18529 - 0.77842i)$<br>$\pm 1.89683i$ | $-1.66947$ |
| $L_6$ | $+0.72864$ | 0 | 0 | $\pm(1.09215 + 0.72243i)$<br>$\pm(1.09215 - 0.72243i)$<br>$\pm 1.82805i$ | $-1.29642$ |
| $L_7$ | $-0.72864$ | 0 | 0 | $\pm(0.96796 + 0.99577i)$<br>$\pm(0.96796 - 0.99577i)$<br>$\pm 1.37504i$ | $-1.29642$ |

The displacement of the equilibrium configurations in the limit as $R \to 0$ is shown in Fig. 4. It can be seen that the collinear points $L_1$, $L_2$, and $L_3$ and triangular points $L_4$ and $L_5$ are displaced towards the equivalent classical Lagrange points. These are denoted as $\bar{L}_i$ ($i = 1 - 5$) in Fig. 4, and are recovered when $R = 0$. However, the two new collinear points $L_6$ and $L_7$ are displaced towards the two primary masses as the ring radius shrinks with a singularity occurring in the limit as $R \to 0$.

Moreover, for $\mu \ll 1$ a number approximations can be made for the location of the collinear equilibria. First, if $< R_H$, where $R_H = (\mu/3)^{1/3}$ is the Hill radius, then the sequence of collinear points $(L_3, L_7, L_1, L_6, L_2)$ can be approximated by $(-1, 0, -R_H, 1, R_H)$. Here, the ring can be seen as point-like, while it may still enclose $m_1$ or $m_2$. However, if $R > R_H$, the sequence can be approximated by $(-1, 0, 1 - R, 1, 1 + R)$, since the collision set $\Sigma_2$ dominates in the vicinity of $m_2$ and so the equilibria are displaced to the edge of $\Sigma_2$, as defined by the ring radius.

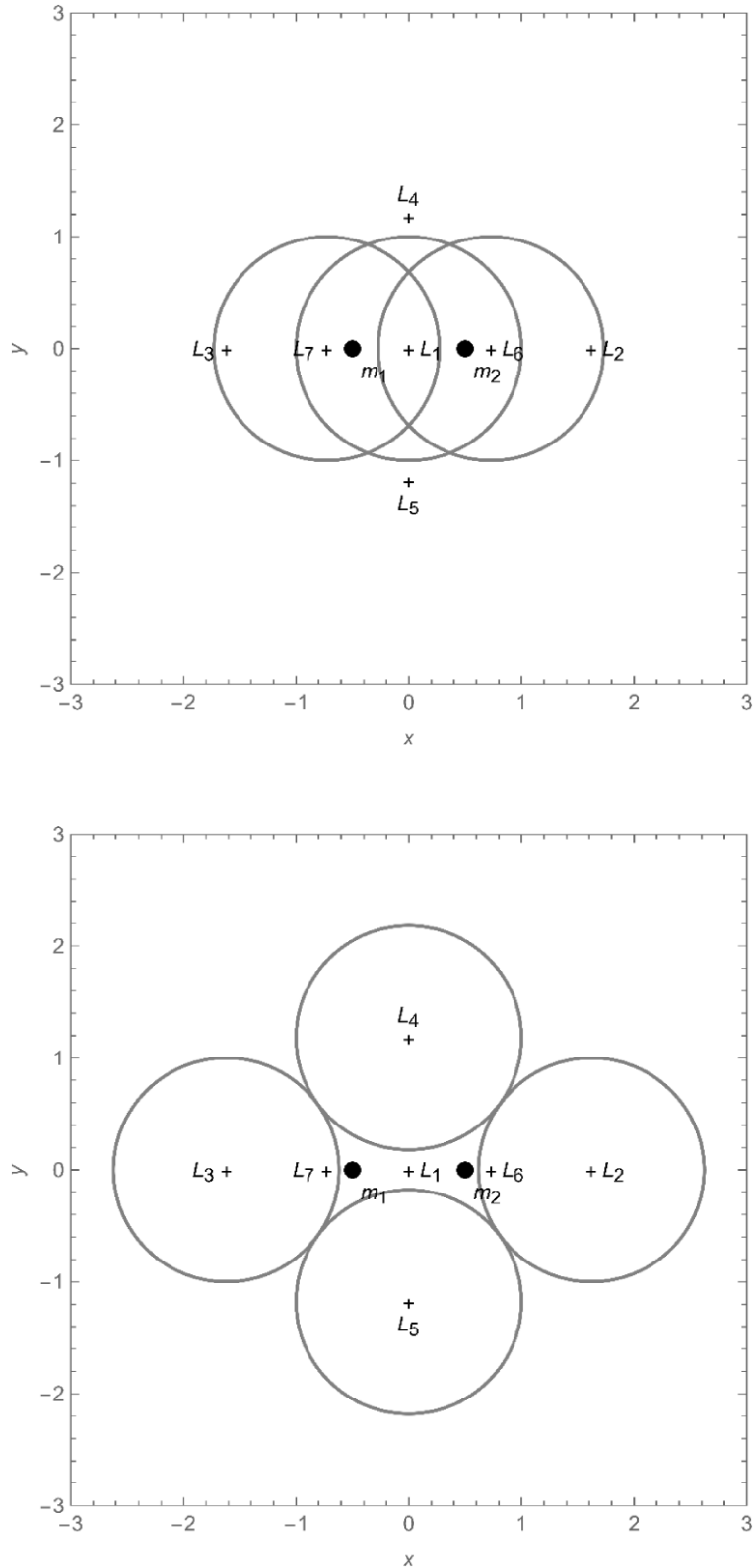

**Figure 3.** Equilibrium configurations for (a) $L_1$, $L_6$, and $L_7$ and (b) $L_2$, $L_3$, $L_4$, and $L_5$ for the equal mass problem with $\mu = 1/2$ and unit ring radius $R = 1$.

Lastly, while the current paper considers a planar ring, it can be shown that a vertical ring can also generate equilibria. Consider the equal mass problem with $\mu = 1/2$ and a vertical ring, where the ring normal is fixed in the rotating frame of reference and is directed along the line connecting $m_1$ and $m_2$. It can be demonstrated, and indeed seen intuitively by symmetry, that if the centre of the ring is located at the origin the ring will be in equilibrium. Since the ring normal is fixed in the rotating frame the ring normal will rotate continuously in an inertial frame. Other equilibria can also be found. Again for the equal mass problem, and now with $R = 0.5$, a set of equilibria are found along the line connecting $m_1$ and $m_2$ at points ($L_3$, $L_7$, $L_1$, $L_6$, $L_2$) given by ($-0.94$, $-0.56$, $0$, $+0.56$, $+0.94$). These correspond to the centre of the ring located at the origin ($L_1$), as noted above, the centre of the ring located close to each of the primary masses ($L_{6/7}$) and two distant equilibria ($L_{2/3}$). Although not discussed further here, off-axis equilibria can also be found.

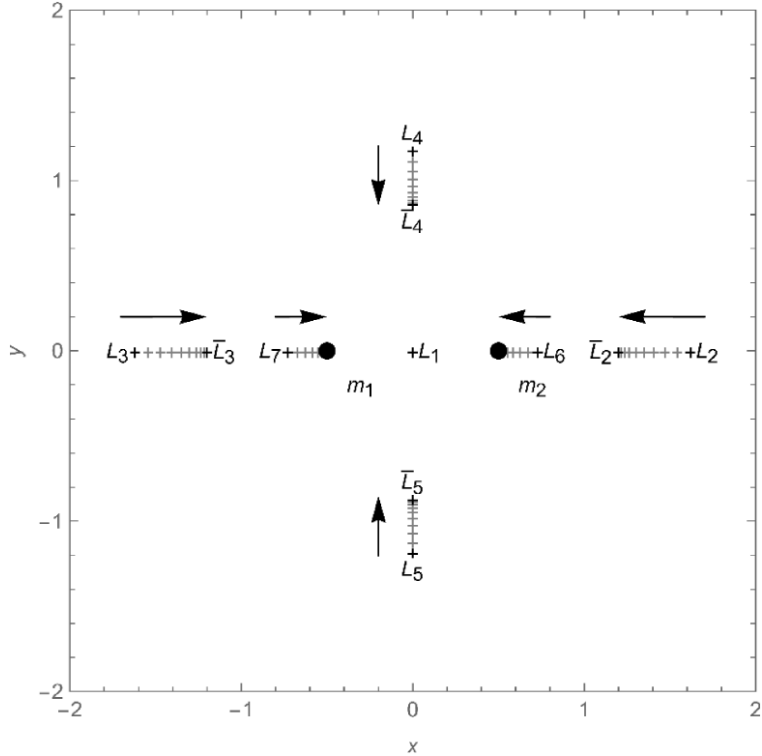

**Figure 4.** Displacement of the location of the equilibrium points as $R \to 0$ for the equal mass problem with $\mu = 1/2$ and initial unit ring radius $R = 1$. Classical Lagrange points are denoted as $\bar{L}_i$ ($i = 1 - 5$).

## 4. EFFECTIVE POTENTIAL AND JACOBI INTEGRAL

Now that the equilibrium configurations of the ring-restricted three-body problem have been identified, the effective potential of the problem $V$ can now be considered where:

$$V = -\frac{1}{2}|\boldsymbol{\omega} \times \boldsymbol{r}|^2 - \frac{2(1-\mu)}{\pi R}\left[\frac{1}{k\sqrt{1-\gamma}}K(\Lambda)\right]_{\bar{r}=\bar{r}_1}$$
$$- \frac{2\mu}{\pi R}\left[\frac{1}{k\sqrt{1-\gamma}}K(\Lambda)\right]_{\bar{r}=\bar{r}_2} = 0. \quad (7)$$

For illustration, the effective potential of the equal mass ring-restricted three-body problem will again be considered with $\mu = 1/2$, and with unit ring radius $R = 1$, as shown in Fig. 5. The location of the seven equilibria can be seen, along with the collision sets defined by equation (5). It can be noted that the coordinates used to define the effective potential correspond to the location of the centre of the ring $C$. Therefore, the collision set in the effective potential shown in Fig. 5 again corresponds to the set of points where, if the ring were placed there, would contact a primary mass resulting in a singularity.

Since the ring-restricted three-body problem is conservative, a Jacobi-type integral can be defined. Following the usual procedure for the restricted three-body problem, the scalar product of equation (4) with the ring velocity vector can be written as

$$\dot{\boldsymbol{r}} \cdot \ddot{\boldsymbol{r}} + 2\dot{\boldsymbol{r}} \cdot (\boldsymbol{\omega} \times \dot{\boldsymbol{r}}) + \dot{\boldsymbol{r}} \cdot \nabla V = 0. \quad (8)$$

The term $2\dot{\boldsymbol{r}}.(\boldsymbol{\omega} \times \dot{\boldsymbol{r}})$ can then be re-written using the scalar triple product identify such that $2\dot{\boldsymbol{r}} \cdot (\boldsymbol{\omega} \times \dot{\boldsymbol{r}}) = 2\boldsymbol{\omega} \cdot (\dot{\boldsymbol{r}} \times \dot{\boldsymbol{r}})$, which vanishes. It can be seen that equation (8) can now be written as

$$\frac{d}{dt}\left[\frac{1}{2}\dot{\boldsymbol{r}} \cdot \dot{\boldsymbol{r}} + V\right] = 0. \quad (9)$$

In the usual manner, a Jacobi-type integral is therefore defined by integrating equation (9) so that:

$$\dot{\boldsymbol{r}} \cdot \dot{\boldsymbol{r}} + 2V = C, \quad (10)$$

where $C$ is the Jacobi constant of the problem. The region of allowed motion is defined by $\dot{\boldsymbol{r}} \cdot \dot{\boldsymbol{r}} > 0$ and so $C - 2V > 0$, and the forbidden region is defined by where $\dot{\boldsymbol{r}} \cdot \dot{\boldsymbol{r}} < 0$ and so $C - 2V < 0$. The family of zero velocity curves are then defined by $\dot{\boldsymbol{r}} = 0$ and so are parametrized by $C$, such that $C - 2V = 0$. For illustration, the zero velocity curves for the equal mass ring-restricted three-body problem will be considered with $\mu = 1/2$, and unit ring radius $R = 1$. The zero velocity curve is shown in Fig. 6 with the Jacobi constant evaluated at the $L_2$ and $L_3$ points where $C = -4.40565$, with the forbidden region $\dot{\boldsymbol{r}} \cdot \dot{\boldsymbol{r}} < 0$ shown as shaded. The potential $V$ at each of the equilibrium points is listed in Table 1.

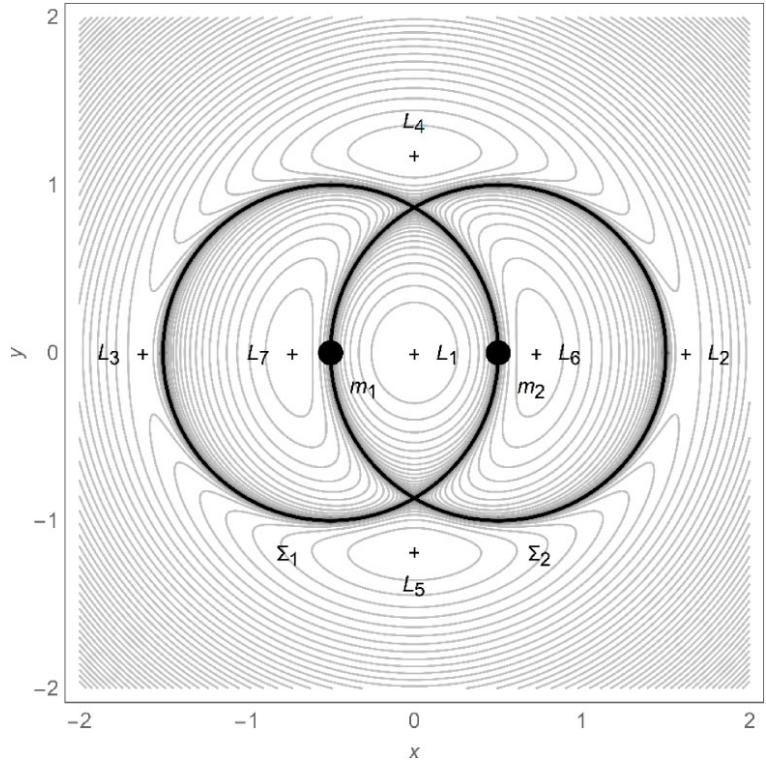

**Figure 5.** Effective potential $V$ for the equal mass problem with $\mu = 1/2$ and unit ring radius $R = 1$.

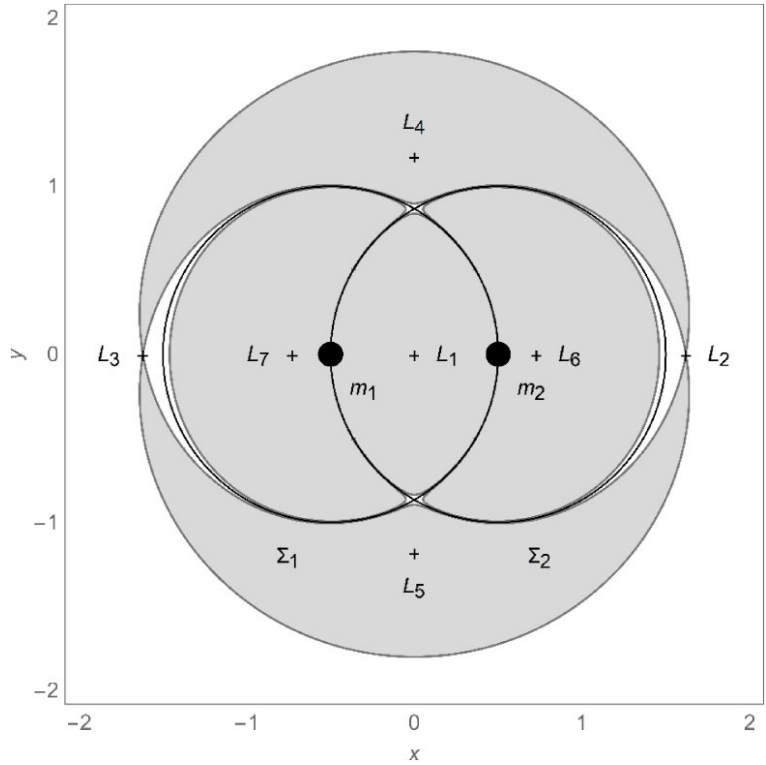

**Figure 6.** Zero velocity curve for the equal mass problem with $\mu = 1/2$ and unit ring radius $R = 1$. The Jacobi constant associated with the $L_2$ and $L_3$ points is shown such that $C = -4.40565$.

## 5. STABILITY PROPERTIES OF THE EQUILIBRIA

In order to determine the stability properties of the equilibrium configurations found in Section 4, the dynamics of the ring-restricted three-body problem will now be linearized. Then, an arbitrary equilibrium solution $\tilde{\boldsymbol{r}}$ can be defined and an infinitesimal perturbation $\delta \boldsymbol{r}$ added such that $\boldsymbol{r} \rightarrow \tilde{\boldsymbol{r}} + \delta \boldsymbol{r}$. From equation (4), the resulting linearized dynamics are defined as

$$\delta \ddot{\boldsymbol{r}} + 2 \boldsymbol{\omega} \times \delta \dot{\boldsymbol{r}} + \left[\frac{\partial}{\partial \boldsymbol{r}} \nabla V\right]_{\boldsymbol{r}=\tilde{\boldsymbol{r}}} \delta \boldsymbol{r} = 0 \qquad (11)$$

which can be written more compactly as

$$\delta \ddot{\boldsymbol{r}} + \Gamma_1 \delta \dot{\boldsymbol{r}} + \Gamma_2 \delta \boldsymbol{r} = 0. \qquad (12)$$

where the skew-symmetric gyroscopic matrix $\Gamma_1$ and Hessian matrix $\Gamma_2$ are defined as

$$\Gamma_1 = \begin{bmatrix} 0 & -2 & 0 \\ 2 & 0 & 0 \\ 0 & 0 & 0 \end{bmatrix} \qquad (13a)$$

$$\Gamma_2 = \left[\frac{\partial}{\partial \boldsymbol{r}} \nabla V\right]_{\boldsymbol{r}=\tilde{\boldsymbol{r}}}. \qquad (13b)$$

In order to proceed, a trial solution can be substituted into equation (11) of the form $\delta \boldsymbol{r}(t) = \delta \bar{\boldsymbol{r}} e^{\lambda t}$ for some constant vector $\delta \bar{\boldsymbol{r}}$ and exponent $\lambda$. This leads to the characteristic polynomial of the problem which can be found from:

$$\det \|\lambda^2 \boldsymbol{I} + \lambda \Gamma_1 + \Gamma_2 \| = 0. \qquad (14)$$

The stability properties of the collinear and triangular points will now be investigated. Using equation (14) the eigenvalues associated with each equilibrium point can be found as a function of both the mass ratio of the problem $\mu$ and the ring radius $R$. As an extension to the classical restricted three-body problem, the ring radius provides an additional free parameter that generates new stability characteristics for the ring-restricted three-body problem.

### 5.1 Collinear points

For the collinear points $y = z = 0$, it is found that the off-diagonal terms in equation (13b) vanish such that the Hessian matrix is given by

$$\Gamma_2 = \begin{bmatrix} a & 0 & 0 \\ 0 & b & 0 \\ 0 & 0 & c \end{bmatrix}, \qquad (15)$$

where $a = V_{xx}|_{\boldsymbol{r}=\tilde{\boldsymbol{r}}}$, $b = V_{yy}|_{\boldsymbol{r}=\tilde{\boldsymbol{r}}}$, and $c = V_{zz}|_{\boldsymbol{r}=\tilde{\boldsymbol{r}}}$. Then, the characteristic polynomial defined by equation (14) can be written as

$$\left(\lambda^2 + c\right)\left(\lambda^4 + \lambda^2(a+b+4) + ab\right) = 0 \qquad (16)$$

which has solutions:

$$\lambda_{1,2} = \pm \frac{\sqrt{-4-a-b+\sqrt{-4ab+(4+a+b)^2}}}{\sqrt{2}} \qquad (17a)$$

$$\lambda_{3,4} = \pm \frac{\sqrt{-4-a-b-\sqrt{-4ab+(4+a+b)^2}}}{\sqrt{2}} \qquad (17b)$$

$$\lambda_{5,6} = \pm i \sqrt{c}. \qquad (17c)$$

In general, determining the nature of the eigenvalues is complicated by the functional form of the terms $a$, $b$, and $c$ representing the partial derivatives of the potential. However, as noted by Binney and Tremaine (1988), if $\lambda$ is a solution of equation (16) then $-\lambda$ is also a solution. Therefore, if $\lambda$ has a real part, there will be a counterpart eigenvalue with a real part of opposite sign. Any eigenvalue with a real part will therefore lead to instability. For stability, purely imaginary eigenvalues are therefore required, and so $\lambda^2 < 0$. First, it can be seen from equation (16) that linear stability requires $c > 0$. Moreover, the biquadratic term in equation (16) will yield $\lambda^2 < 0$ if its coefficients are positive, so that $a + b + 4 > 0$ and $ab > 0$, and its discriminant is positive (to ensure that $\lambda^2$ is real) so that $(a+b+4)^2 - 4ab > 0$. For illustration, the equal mass problem can again be considered with $\mu = 1/2$ and unit ring radius $R = 1$. The equilibrium point $L_1$ is then defined by $x = 0$, $y = 0$, and $z = 0$ by symmetry, and it is found that the coefficients $a = -2.20$, $b = -1.69$, and $c = 1.89$. As expected the out-of-plane motion is linearly stable since $c > 0$. Although $a + b + 4 = 0.11$ and $ab = 3.72$, the discriminant $(a+b+4)^2 - 4ab = -14.86$, and so the $L_1$ point in unstable.

Now, in order to investigate the general stability properties of the collinear equilibrium points for an arbitrary mass ratio and ring radius, the eigenvalues defined by equation (17) can be calculated numerically on a grid in the $\mu - R$ parameter space. For each combination $(\mu, R)$, the location of the associated equilibrium point is firstly determined before the eigenvalues are calculated. Each point on the grid can then be designated as linearly stable or unstable. For illustration, only the $L_6$ point will now be considered as shown in Fig. 7. It can be seen that linear stability is possible for a small mass ratio with $\mu < \mu_M$ where $\mu_M = 0.0044$ (and $R = 0.464$) and within a limited range of ring radii $R$. As the mass ratio of the problem decreases the range of stable ring radii grows, with a maximum stable ring radius of $R_M = 0.705$ in the limit that $\mu \rightarrow 0$. This corresponds to a ring-restricted two-body problem, which will be discussed later in Appendix B. In the limit that both $\mu \rightarrow 0$ and $R \rightarrow 0$, the two-body problem with a point mass is recovered. The implications of the narrow region of stability in the $\mu - R$ parameter space will be discussed later in Section 7. However, it can be seen that a ring enclosing the smaller of the primary masses can in principle be linearly stable. An example stable response of a ring in the vicinity of the $L_6$ point is shown in Fig. 8.

### 5.2 Triangular points

In the limiting case when $R \rightarrow 0$, the ring-restricted three-body problem will reduce to the classical circular restricted three-body problem. It is well known that the triangular equilibrium points in the restricted three-body problem can be linearly stable (Battin, 1999), provided the mass ratio of the problem is less than Routh's value $\mu_R = (1/2) - \sqrt{23/108}$. It can therefore be expected that the ring-restricted three-body problem will exhibit a change in stability properties when $R \ll 1$ and $\mu \sim \mu_R$. Again, in order to investigate the stability properties of the ring-restricted three-body problem at the triangular equilibrium points, the eigenvalues can be determined from equation (14) on a grid in the vicinity of Routh's value in the $\mu - R$ parameter space. For each combination $(\mu, R)$, the location of the associated equilibrium point is determined before the eigenvalues are calculated. Direct calculation from equation (14) is now required since the mixed partial derivatives of the potential may now be non-zero. Each point on the grid can again be designated as linearly stable or unstable, as shown in Fig. 9.

It can be seen that the ring-restricted three-body problem is always be unstable for $\mu > \mu_R$. However, for $\mu < \mu_R$ there is a linearly stable region for $R \neq 0$. Again, as the mass ratio of the problem decreases the range of stable ring radii grows, with a maximum stable ring radius of $R_M = 0.705$ in the limit that $\mu \to 0$. This corresponds to a ring-restricted two-body problem, as discussed for the collinear case. The implications of this region of stability in the $\mu - R$ parameter space will again be discussed later in Section 7. An example stable response of a ring in the vicinity of the $L_4$ point is shown in Fig. 10. The eigenvalues at each of the equilibrium points are listed in Table 1 for the equal mass case with $\mu = 1/2$, and with unit ring radius $R = 1$.

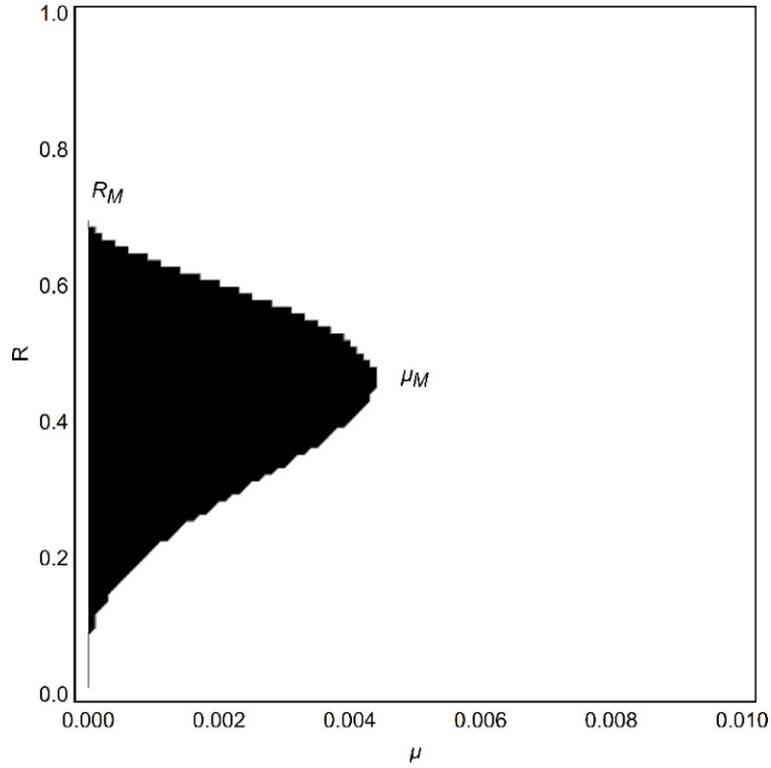

**Figure 7.** Stability region (shaded) for the collinear $L_6$ point as a function of $\mu$ and $R$.

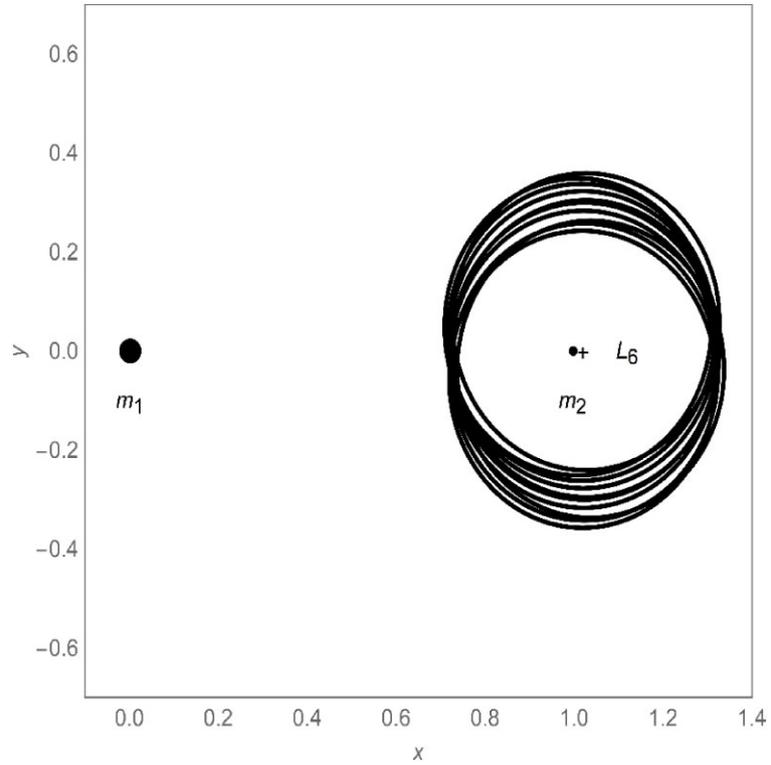

**Figure 8.** Stable response of a ring of radius $R = 0.3$ at the $L_6$ point with $\mu = 2 \times 10^{-3}$ and initial displacement $\delta \boldsymbol{r} = (10^{-3}, 10^{-3}, 10^{-3})$ (for non-dimensional integration time $t_f = 100$).

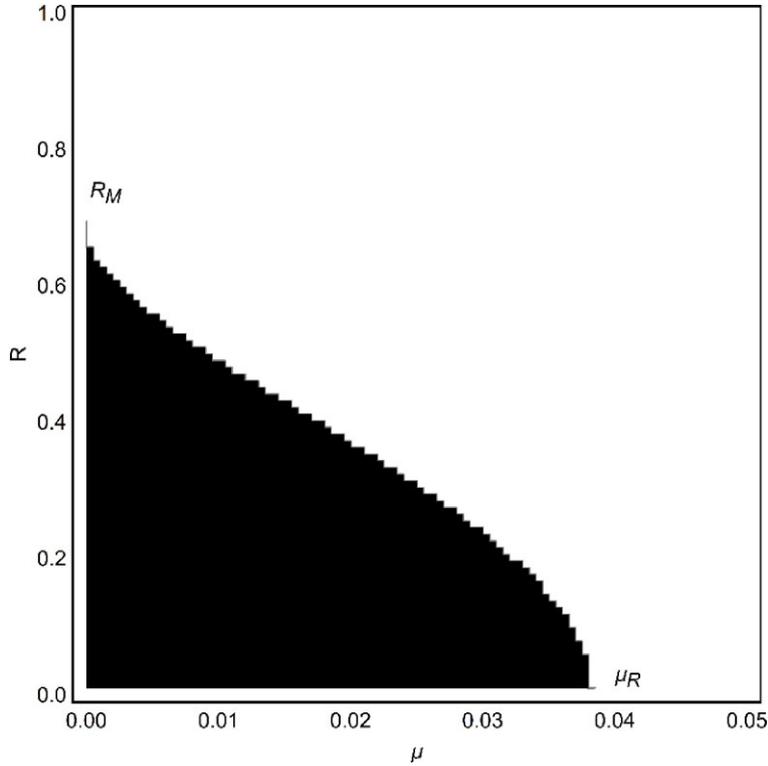

**Figure 9.** Stability region (shaded) for the triangular $L_{4/5}$ points as a function of $\mu$ and $R$.

## 6. SHELL-RESTRICTED THREE-BODY PROBLEM

As an extension to the ring-restricted three-body problem, consider now the case where the 1D infinitesimal ring, embedded in $\mathbb{R}^2$, is replaced by a 2D infinitesimal hollow shell, embedded in $\mathbb{R}^3$. It has been shown in Section 4 that the topology of the ring leads to two new equilibria, so it is of interest to extend the analysis from a closed ring to a closed shell. Indeed, the shell and ring-restricted three-body problems are related in the geometric sense that a ring is a section of a shell. Moreover, if the centre of the ring is superimposed on a central mass the gravitational interaction of the ring with the mass it encloses vanishes due to symmetry. It is therefore anticipated that a similar set of equilibria will be found for the shell-restricted three-body problem. However, since the shell forms a closed surface, Gauss' theorem can be invoked which leads directly to Newton's shell theorem. Herein lies the difference between the two problems. For the closed shell there is no nett gravitational interaction with the mass that it encloses, even when the symmetry is broken and the centre of the shell is displaced. However, for the ring there is an interaction force after the symmetry is broken since it is not fully closed.

In order to proceed, Newton's shell theorem (Reed, 2022) can be invoked to provide insights into the problem. First, it can be noted that the shell theorem states that for a mass outside of the shell, the shell interacts with the mass gravitationally as if the mass of the shell were concentrated at a point at its centre. Furthermore, if the shell encloses a mass, then no net gravitational force is exerted on the shell by that mass, independent of the location of the mass within the shell. Three cases can now be defined; when the shell encloses (1) both primary masses, (2) neither of the primary masses, or (3) one primary mass. These cases can now be considered in turn.

*Case 1 (shell encloses both primary masses)*: Newton's shell theorem states that there will be no nett gravitational force acting on the shell due to the two primary masses which the shell encloses. Therefore, the primary masses form a two-body problem, which is decoupled from the shell. The centre of the shell $C$ can therefore be placed at rest at the centre-of-mass of the two primary masses, or indeed any location where the two primary masses are always enclosed by the shell. Since the shell is gravitationally decoupled from the primary masses, the equilibrium configuration is marginally unstable since the shell is free to drift relative to the primary masses (and indeed collide with them).

*Case 2 (shell encloses neither of the primary masses)*: Newton's shell theorem states that the shell appears gravitationally as a point mass to both of the primary masses. This case is therefore equivalent to the classical restricted three-body problem. The shell can therefore be located at any of the five classical equilibrium points, such that the shell does not enclose either or both of the primary masses. The collinear equilibria are then always unstable and the stability properties of the triangular equilibria are determined by the mass ratio of the problem via Routh's value.

*Case 3 (shell encloses one primary mass)*: Newton's shell theorem again states that there is no net gravitational force acting on the shell due to the single primary mass it encloses. Moreover, the shell will appear gravitationally as a point mass to the other primary mass.

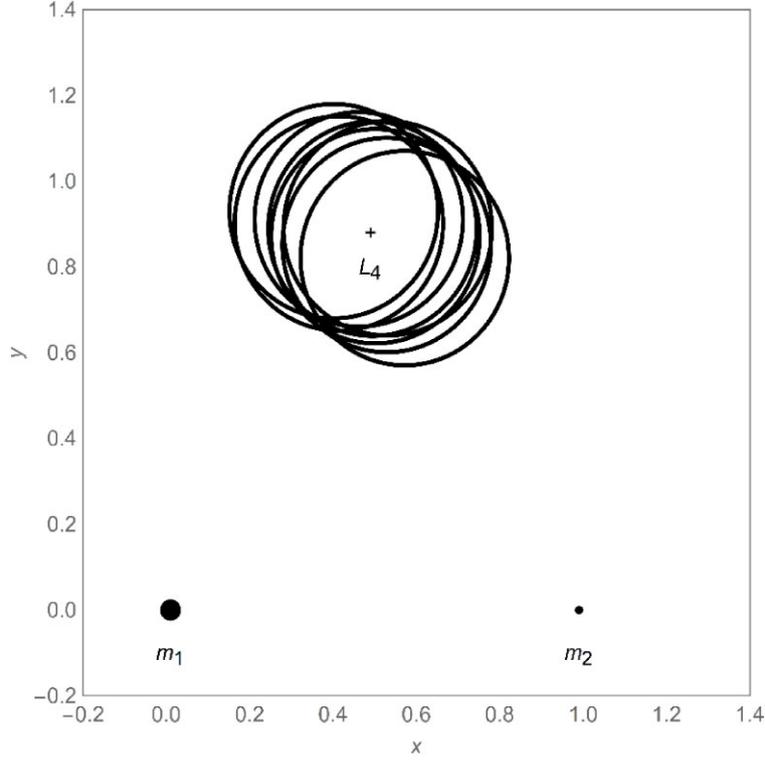

**Figure 10.** Stable response of a ring of radius $R = 0.25$ at the $L_4$ point with $\mu = 10^{-2}$ and initial displacement $\delta r = (3 \times 10^{-2}, 3 \times 10^{-2}, 3 \times 10^{-2})$ (non-dimensional integration time $t_f = 100$).

This is an interesting case since the primary mass that the shell does not enclose will orbit around the common centre-of-mass of both primary masses. However, the shell will only experience the gravitational potential of the mass that it does not enclose, as if the shell were a point mass. From equation (4), the equation of motion of the shell in the rotating frame of reference for Case 3 can be written as

$$\ddot{r} + 2\omega \times \dot{r} + \nabla V_S(\epsilon) = 0, \tag{18}$$

where $\omega$ is again the angular velocity of the rotating frame of reference, defined by the two primary masses $m_1$ and $m_2$, and the centre of the shell $C$ is located at position $r$. Then, from equation (7) the potential $V_S(\epsilon)$ experienced by the shell can be defined by

$$V_S(\epsilon) = -\frac{1}{2}|\omega \times r|^2 - \epsilon\frac{1-\mu}{r_1} - (1-\epsilon)\frac{\mu}{r_2}, \ \epsilon = \begin{cases} 0, \\ 1 \end{cases} \tag{19}$$

where $\mu = m_2/m_1 + m_2$ is again the mass ratio of the two primary masses. The two gravitational potential terms in equation (19) represent the interaction of the shell with each of the primary masses. The distance from each of the primary masses to the centre of the shell $C$ is given by $r_1 = \|r - \bar{r}_1\|$ and $r_2 = \|r - \bar{r}_2\|$, where again $\bar{r}_1 = (-\mu, 0, 0)$ and $\bar{r}_2 = (1-\mu, 0, 0)$. The index $\epsilon$ indicates which of the primary masses the shell encloses, with $\epsilon = 0$ indicating that the shell encloses primary mass $m_1$ and $\epsilon = 1$ indicating that the shell encloses primary mass $m_2$, where it is assumed that $m_1 > m_2$.

Extending the analysis of the ring-restricted three-body problem, collision sets $\Sigma_1$ and $\Sigma_2$ can be defined as the intersection of the shell of non-dimensional radius $R$ with either of the primary masses $m_1$ or $m_2$ such that:

$$\Sigma_1 = \left(r \in \mathbb{R}^3 : (x+\mu)^2 + y^2 + z^2 = R^2\right) \tag{20a}$$

$$\Sigma_2 = \left(r \in \mathbb{R}^3 : (x-(1-\mu))^2 + y^2 + z^2 = R^2\right) \tag{20b}$$

as shown in Fig. 11, corresponding to the equal mass shell-restricted three-body problem with $\mu = 1/2$ and with ring radius $R = 0.45$. The shell-restricted three-body problem is therefore defined on $\mathbb{R}^3 - (\Sigma_1 \cup \Sigma_2)$. Again, since the infinitesimal mass of the shell-restricted three-body problem is extended, rather than a point mass, the singularities of the problem are represented by spheres. However, since the centre-of-mass of the shell is empty, it can encompass one or both of the primary masses so that there is no singularity when the centre of the shell is coincident with a primary masses (unless the radius of shell is equal to the separation of the primary masses).

From equation (18) the condition for an equilibrium solution at some position $\tilde{r}$ with $\ddot{r} = \dot{r} = 0$ is given by $\nabla V_S(\epsilon) = 0$, where there are two branches of solutions corresponding to $\epsilon = 0$ and $\epsilon = 1$. It can then be shown from equation (19) that the equilibrium solution is collinear with $\nabla V \cdot e_2 = 0$ and $\nabla V \cdot e_3 = 0$ when $y = z = 0$. Moreover, it can also be shown from equation (19) that the further condition required for equilibrium that $\nabla V \cdot e_1 = 0$ is obtained when:

$$-x + \frac{\mu(x-(1-\mu))}{\|x-(1-\mu)\|^3} = 0, \ \epsilon = 0 \tag{21a}$$

$$-x + \frac{(1-\mu)(x+\mu)}{\|x+\mu\|^3} = 0, \ \epsilon = 1. \tag{21b}$$

It can be seen that the equilibrium configuration represented by equation (21a), with $\epsilon = 0$ indicating that the shell encloses primary mass $m_1$, is defined by $\tilde{r} = (-\mu, 0, 0)$. It can also be shown that there exists an additional solution with $x > 0$; however, the solution is not physical since for $\epsilon = 0$ the shell must enclose the larger primary mass. Similarly, it can be seen that the equilibrium solution represented equation (21b), with $\epsilon = 1$ indicating that the shell encloses primary mass $m_2$, is defined by $\tilde{r} = (1-\mu, 0, 0)$. Again, it can also be shown that there exists an additional solution with $x < 0$; however, the solution is not physical since for $\epsilon = 1$ the shell must enclose the smaller primary mass.

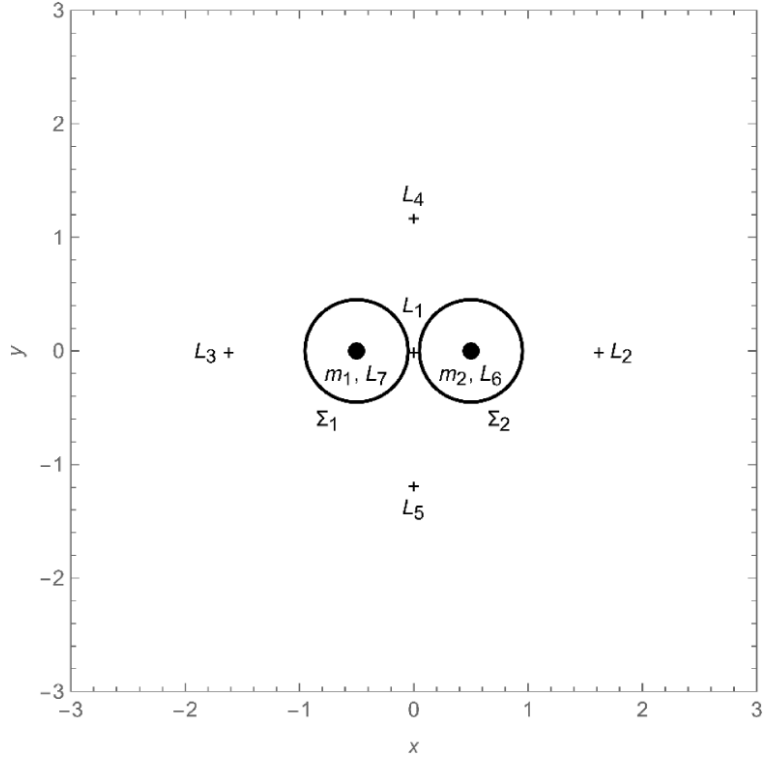

**Figure 11.** Location of the seven equilibrium points and collision sets $\Sigma_1$ and $\Sigma_2$ for the equal mass case with $\mu = 1/2$ and shell radius $R = 0.45$.

For an equilibrium configuration, the centre of the shell $C$ is therefore collocated with the primary mass which it encloses, resulting in two new equilibria, denoted $L_6$ and $L_7$ as shown in Fig. 11. These two new equilibria appear since the interior of the shell is empty and so its centre $C$ can be superimposed over the primary masses $m_1$ and $m_2$ without resulting in a singularity. These results are independent of the radius of the shell, due to Newton's shell theorem, unlike the ring-restricted three-body problem where the location of the equilibria is a function of the ring radius.

In order to illustrate the configuration of the shell and primary masses, the shell is superimposed on the seven equilibria in Fig. 12, corresponding to cases 2 and 3 above. It can be seen that at the $L_1$ point the shell is located between the primary masses, with the diameter of the shell less than the spacing between the masses. At the $L_6$ and $L_7$ points the shell encloses one of the primary masses, while at the $L_2$, $L_3$, $L_4$, and $L_5$ points the shell does not enclose either of the primary masses.

While the shell does not interact directly with the mass it is enclosing due to Newton's shell theorem, the dynamics of the shell are still indirectly coupled to both of the primary masses. For example, in the case $\epsilon = 1$ the shell encloses primary mass $m_2$, and so the shell only interacts directly with primary mass $m_1$. However, the location of $m_2$ is defined by the motion of both primary masses about their common centre-of-mass. Therefore, while the shell only interacts directly with mass $m_1$ in a two-body gravitational interaction, its motion is highly non-Keplerian. This is due to the gravitational interaction of $m_1$ and $m_2$ which defines the rotating frame of reference, the location of the primary masses and hence the location of the equilibria.

In order to investigate the linear stability properties of the new equilibria equation (18) can now be linearized. The equilibrium solutions $\tilde{r}$ can be defined where $\tilde{r} = (-\mu, 0, 0)$ when $\epsilon = 0$ corresponding to $L_7$, and where $\tilde{r} = (1 - \mu, 0, 0)$ when $\epsilon = 1$ corresponding to $L_6$. An infinitesimal perturbation $\delta r$ is again applied such that $r \to \tilde{r} + \delta r$. From equation (18), the resulting linearized dynamics can be written as

$$\delta \ddot{r} + 2\omega \times \delta \dot{r} + \left[\frac{\partial}{\partial r} \nabla V_S(\epsilon)\right]_{r=\tilde{r}} \delta r = 0, \quad \epsilon = \begin{cases} 0 \\ 1 \end{cases}. \quad (22)$$

The collinear analysis from Section 5 can now be used to determine the eigenvalues of the problem using equation (17), but with the new potential such that $a = V_S(\epsilon)_{xx}|_{r=\tilde{r}}$, $b = V_S(\epsilon)_{yy}|_{r=\tilde{r}}$, and $c = V_S(\epsilon)_{zz}|_{r=\tilde{r}}$, where the mixed derivatives vanish at the collinear equilibria. Again, there will be two branches of solutions corresponding to $\epsilon = 0$ and $\epsilon = 1$. From equation (19) it can be shown that for $\epsilon = 1$, corresponding to the $L_6$ point, $a = -3 + 2\mu$, $b = -\mu$, and $c = 1 - \mu$, while for $\epsilon = 0$, corresponding to the $L_7$ point, $a = -(1 + 2\mu)$, $b = -1 + \mu$, and $c = \mu$.

First, it can be noted in both branches $c > 0$ for $0 < \mu \le 1/2$ and so from equation (17c) the out-of-plane motion is linearly stable. Then, to investigate the conditions for in-plane linear stability, the biquadratic generated by equation (16) can be written as

$$\sigma^2 + \sigma(a + b + 4) + ab = 0, \quad (23)$$

where $\sigma = \lambda^2$, where both roots of equation (23) must be real and negative, as discussed in Section 5. The coefficients of the quadratic can be determined and it can be seen that for $0 < \mu \le 1/2$:

$$a + b + 4 = 1 + \mu > 0 \quad (24a)$$

$$ab = \mu(3 - 2\mu) > 0 \quad (24b)$$

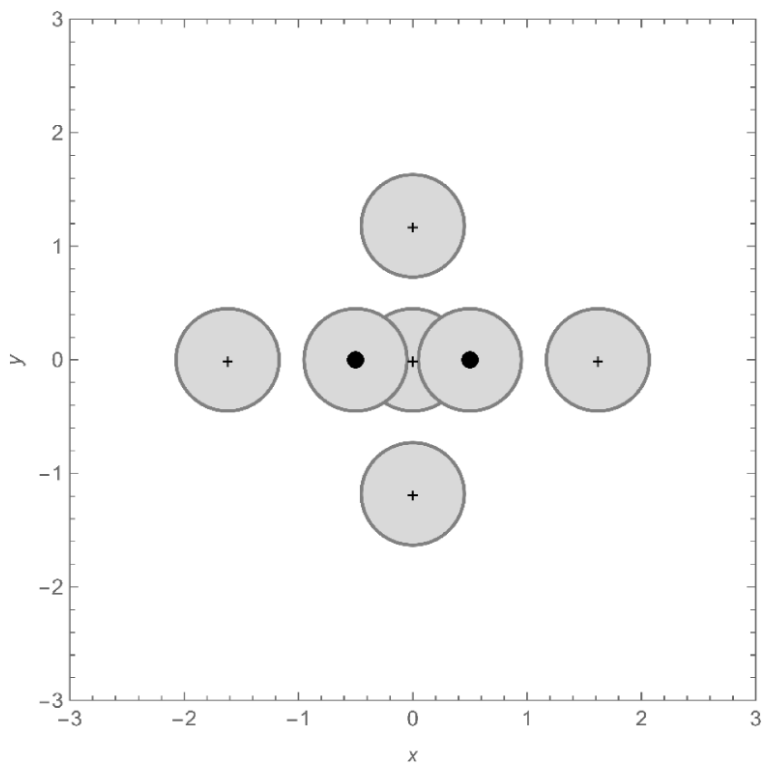

**Figure 12.** Set of seven shell equilibrium configurations for the equal mass case with $\mu = 1/2$ and shell radius $R = 0.45$.

and so both coefficients are positive as a necessary condition for negative roots. For both roots to be real, the discriminant of the biquadratic defined by equation (16) must also be positive so that $(a + b + 4)^2 - 4ab > 0$, which leads to a quadratic in the mass ratio such that:

$$9\mu^2 - 10\mu + 1 > 0 \qquad (25)$$

and so $(9\mu - 1)(\mu - 10) > 0$. The inequality in equation (25) is therefore satisfied if $9\mu - 1 > 0$ and $\mu - 10 > 0$ or $9\mu - 1 < 0$ and $\mu - 10 < 0$. However, since $0 < \mu \leq 1/2$ it can be seen that the inequality is satisfied only when $\mu < \mu_C$ where $\mu_C = 1/9$. The collinear point $L_6$ can therefore be stable. It can also be shown that the collinear point $L_7$ with is always unstable. Therefore, when the shell is collocated with the larger of two masses at $\tilde{r} = (-\mu, 0, 0)$ the equilibrium point is always unstable, while if the shell is collocated with smaller of two masses at $\tilde{r} = (1 - \mu, 0, 0)$ the equilibrium point can be linearly stable if $\mu < \mu_C$. The transition from stability to instability for the $L_6$ point as can be seen in Fig. 13, where a pair of real eigenvalues are created for $\mu > \mu_C$. It can be noted that the stability properties of the new equilibria are independent of the radius of the shell, due to Newton's shell theorem, unlike the earlier analysis presented for the ring-restricted three-body problem.

In the limit that $\mu \to 0$ the eigenvalue spectrum $(\lambda_1, \lambda_2, \lambda_3, \lambda_4, \lambda_5, \lambda_6)$ can be obtained from the collinear stability analysis using equation (17) and is found to be $(0, 0, +i, -i, +i, -i)$, as can also be seen in Fig. 13. The limit $\mu \to 0$ corresponds to the shell orbiting a single mass in what is now a two-body problem, since the shell is gravitationally equivalent to a point mass due to Newton's shell theorem. The repeat zero eigenvalues then correspond to the free azimuthal draft of the shell.

It can be noted that the two new equilibrium configurations in Case 3 exist in addition to the five classical equilibria in Case 2. Therefore, if the radius of the shell is less than the separation of the primary masses, and so the shell can enclose either of the primary masses, the shell-restricted three-body problem has seven equilibria, again shown in Fig. 12, as does the ring-restricted three-body problem. One of the two new collinear equilibria associated with the shell-restricted three-body problem can also be stable. The implications of this finding are discussed in Section 7.

## 7. DISCUSSION

It has been demonstrated that the ring-restricted three-body problem possesses seven equilibrium configurations, two triangular points, and five collinear points. For small mass ratios, the triangular points can be stable (as an extension of Routh's criterion) for a range of ring radii. The $L_6$ point, where the ring encloses the smaller of the primary masses, has been investigated and it has been shown to be stable for small mass ratios and a range of ring radii. It has also been demonstrated that the shell-restricted three-body problem has seven equilibrium configurations, with a stable configuration possible when the shell encloses the smaller of the primary masses. With the properties of these new equilibrium configurations established, a range of applications can now be considered.

### 7.1 Relation to Maxwell's ring instability

As noted in Section 1, Maxwell demonstrated that a uniform ring enclosing a central mass has a single unstable equilibrium configuration, with the centre of the ring superimposed on the central mass (Maxwell, 1859). However, for the ring-restricted three-body problem it has been shown, for example, that the $L_6$ point can be stable for a small mass ratio $\mu \ll 1$, corresponding to a uniform ring encompassing the secondary mass $m_2$. Physically, the ring is in orbit around the large primary mass $m_1$, with the ring radius less than the stable upper bound $R_M$ for the two-body problem resulting from $\mu \rightarrow 0$. The additional gravitational interaction between the ring and the secondary mass $m_2$ then leads to an equilibrium configuration offset from the secondary mass and defined by the $L_6$ point. While Maxwell demonstrated that a uniform ring enclosing a central mass in the two-body problem is unstable, it has been demonstrated here that a ring enclosing the secondary mass can be linearly stable in the restricted three-body problem.

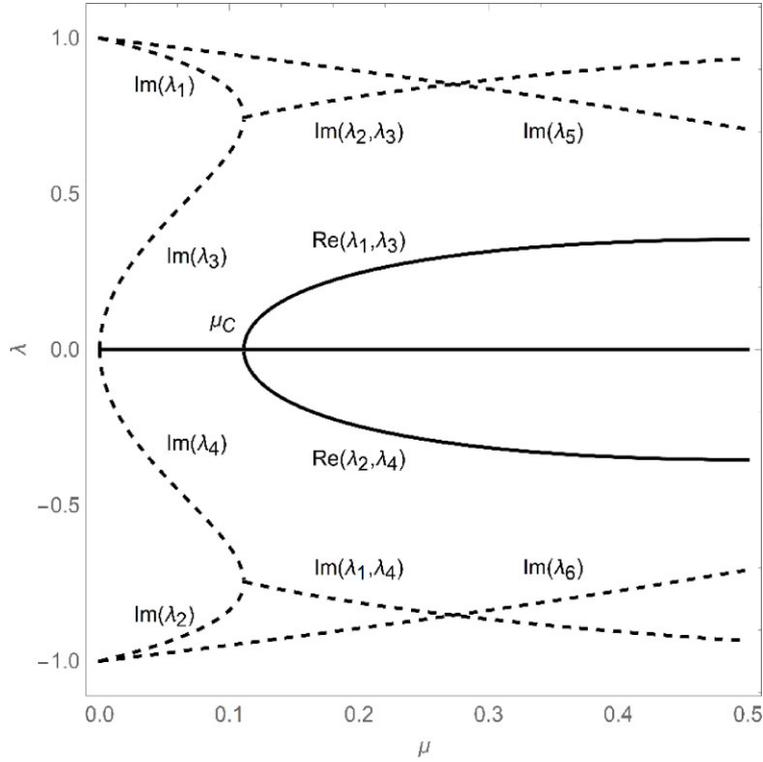

**Figure 13.** Eigenvalues for equilibrium point $L_6$ for the case with $\epsilon = 1$ with the shell enclosing the smaller of the masses $m_2$. Instability occurs when $\mu > \mu_C$ when a pair of real eigenvalues are created (——real part,— — — imaginary part).

In order to illustrate the analysis, the eigenvalues of the $L_6$ point in the Sun–Saturn ring-restricted three-body problem will now be considered with a mass ratio $\mu$ of $2.857 \times 10^{-4}$. In this case $\mu < \mu_M$, the stable upper limit for the mass ratio for the $L_6$ point shown in Fig. 7. The eigenvalues are then determined from equations (17) as a function of the ring radius, where the location of the $L_6$ point must also be determined for each ring radius. This is equivalent to a section through the stability map of Fig. 7 for a fixed mass ratio. It can be seen from Fig. 14 that the real eigenvalues vanish at a lower ring radius $R_L$ of 0.14 and then re-appear at an upper ring radius $R_U$ of 0.67, where $R_U < R_M$ since $\mu > 0$, although in this case $\mu \ll 1$. Therefore, a uniform ring enclosing Saturn can be linearly stable if $0.14 < R < 0.67$, so that the length-scale of such a stable ring is clearly large.

### 7.2 Ringworlds, Dyson spheres, and SETI

A ringworld is envisaged in fiction as a ring enclosing a single star with the centre of the ring superimposed on the star (Niven, 1970). While this configuration is unstable, the analysis presented in Section 5 has determined that a ringworld would in principle be stable in a binary system, if the mass ratio of the problem $\mu < \mu_M$. If it is now assumed, for illustration only, that the densities of the primary masses are similar, then the ratio of the radius of the secondary $R_2$ to the primary $R_1$ would be of the order of $R_2/R_1 \sim 0.16$. A ringworld can therefore be in a stable equilibrium configuration in a binary system if the secondary mass has a radius of order 1/5 of that of the primary mass.

While binary star systems with an extremely low mass ratio such that $\mu < \mu_M$ (for the $L_6$ point) are unlikely, speculatively, a solar mass object with for example a planetary mass brown dwarf (Luhman et al. 2023) partner could in principle possess a small enough mass ratio for ring stability. Moreover, stable artificial rings can in principle be envisaged in star–exoplanet or exoplanet–exomoon systems, while stable rings at the triangular equilibria can also be envisaged for large space habitats. The existence of passively stable orbits for such large-scale structures may have implications for techno-signatures in search for extra-terrestrial intelligence (SETI) studies, although there are significant material limitations for such structures (Fridman et al. 1984).

Moreover, while stability is only possible under certain conditions (small mass ratios and a range of ring radii), the analysis presented in Section 3 has also shown that a range of (unstable) equilibria exist, as shown in Fig. 3. These configurations include a ringworld entirely enclosing a binary star system, orbiting outside of a binary system or enclosing one of the stars.

Similarly, a Dyson sphere is envisaged as a shell enclosing a single star with the centre of the shell superimposed on the star (Dyson, 1960). Due to Newton's shell theorem this location will be marginally unstable. Since there is no net gravitational interaction between the shell and star, any disturbance will result in a uniform drift of the centre of the shell away from the star until a collision occurs. The analysis presented in Section 6 has determined that a Dyson sphere would in principle be stable in a binary star system if the mass ratio of the problem $\mu < \mu_C$. Again for illustration only, if the densities of the primary masses are assumed to be similar then the ratio of the radius of the secondary $R_2$ to the primary $R_1$ is of order $R_2/R_1 \sim 0.46$. A Dyson sphere the can therefore in principle be in a stable equilibrium configuration in a binary system if the secondary mass has a radius of order half that of the primary mass.

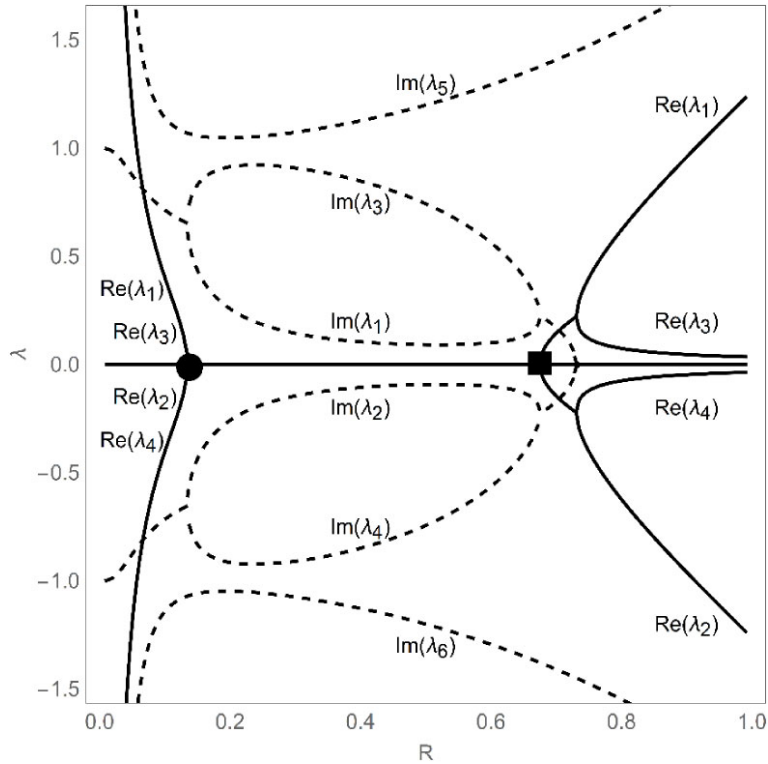

**Figure 14.** Eigenvalues for equilibrium point $L_6$ for the Sun-Saturn system. The stable region is defined by $0.14$ ($R_L$ ●) $< R < 0.67$ ($R_U$ ■) when two pairs of imaginary eigenvalues are created (——real part,---- imaginary part).

Again, the existence of passively stable orbits for such large-scale structures may have implications for techno-signatures in SETI studies. For example, material limitations aside, a strong infra-red excess observed from one member of binary star system with a mass ratio $\mu < \mu_C$ could in principle support the existence of a Dyson sphere. It can be noted that spectroscopic binaries have been considered with a minimal mass ratio $m_2/m_1 \sim 0.08$ (Ducati et al. 2011). Stable shells around star–exoplanet or exoplanet–exomoon systems with a stable mass ratio can in principle also be envisaged. Finally, it can be noted that in addition to a single shell, a Dyson sphere comprised of multiple nested shell can be envisaged (Wright, 2020; Suazo et al. 2022). Since the infinitesimal shells do not interact gravitationally, and due to Newton's shell theorem, the nested shells will also be stable provided the mass ratio $\mu < \mu_C$.

sphere. However, due to geometric imperfections, the mechanics of a thin shell in compression are severely limited by the onset of buckling modes (Wright, 2020). Nevertheless, this assumes that a Dyson sphere is a passive and inert structure. Alternatively, so-called smart materials have been developed with embedded sensors and actuators (Rao and Singh, 2001); for example, elastic shells fabricated with a tuneable buckling strength using active magnetic elastomers (Yan et al. 2021). Given that the speed of sound in graphene will be slow compared to the speed of light, deformations of the shell geometry will propagate significantly slower than feedback signals to actuators across the sphere. It can therefore be speculated that a Dyson sphere could be a smart structure which is able to actively suppress the onset of buckling modes. Moreover, radiation pressure can in principle be used to partly or fully support a thin shell, thus lowering the compressive load on the shell (Wright, 2020). The addition of radiation pressure will clearly modify the orbital dynamics of the problem, but is not considered here.

### 7.3 Material considerations

There are significant material limitations for so-called ringworlds and shell-type Dyson spheres (Fridman et al. 1984; Wright, 2020). For example, the required tensile strength $\sigma$ for a Dyson sphere of radius $R$ in compression about a star of mass $M$ and fabricated from a material of density $\rho$ can be estimated as $\sigma \approx GM\rho/2R$ (Loeb 2023). The quantity $\sigma/\rho$ is then the required specific strength of the material. For a Dyson sphere of radius 1 au centred on a solar mass star $\sigma/\rho = 4 \times 10^5 \, kNm^{-1}$. For graphene, assuming a density close to graphite of $2.26 \times 10^3 \, kgm^{-3}$ and an intrinsic tensile strength of $130 \, GPa$ (Lee et al. 2008), the specific strength is $5.7 \times 10^4 \, kNm^{-1}$, which is within an order of magnitude of the requirement for a Dyson

### 8. CONCLUSIONS

A ring-restricted three-body problem has been defined and the existence and stability properties of new equilibrium solutions investigated. It has been demonstrated that seven equilibrium configurations exist, with five collinear equilibria and two triangular equilibria. It has shown that for a range of ring radii and a narrow range of mass ratios the ring can be in a stable equilibrium configuration while enclosing the smaller of the primary masses, extending Maxwell's findings from the two-body problem. The two triangular equilibria are found to be stable for mass ratios less than Routh's value and for a range of ring radii. The problem has then been extended to a shell-restricted three-body problem, again with new equilibria identified and their stability properties investigated. It has been shown that a shell can be in a stable equilibrium configuration while enclosing the smaller of the primary masses. The existence of passively stable orbits for such large-scale structures may have implications for SETI studies.

**ACKNOWLEDGEMENTS**

This work was supported by the Department of Science, Innovation and Technology (DSIT) and the Royal Academy of Engineering under the Chair in Emerging Technologies programme. For the purpose of open access, the author has applied a Creative Commons Attribution (CC-BY) license to any Author Accepted Manuscript version arising from this submission.

**DATA AVAILABILITY**

The data that support the findings of this study are available from the corresponding author upon reasonable request. Computations were performed using Mathematica 11.3.

**FORCES ACTING A UNIFORM RING IN THE ROTATING FRAME OF REFERENCE**

In order to proceed, the gravitational potential energy due to the interaction of a uniform ring with a point mass will be determined. Previous analysis, such as Lass and Blitzer (1983) and Schumayer and Hutchinson (2019), considers the potential at some arbitrary field point relative to a local coordinate system centred on a fixed ring. However, in the analysis presented here, the location of the centre of the ring is in general free. This more general formulation will be required to evaluate the gravitational potential energy due to the interaction between the ring and the two primary masses in the ring-restricted three-body problem.

**Gravitational force**

First, a frame of reference will be defined with unit vectors ($e_1$, $e_2$, $e_3$) as shown in Fig. A1, with $e_3$ directed normal to the plane. Then, for generality, an infinitesimal ring with domain $\Omega$, radius $R$, and centre $C$ will be located at some arbitrary position $r$ with coordinates $(x, y, z)$, where it will be assumed that the ring normal fixed and is directed along the $e_3$ axis.

A point mass $M$ will now be located at position $\bar{r}$ with coordinates $(\bar{x}, \bar{y}, \bar{z})$, where the vector connecting the point mass to the ring centre is given by $r' = r - \bar{r}$. For clarity, and consistency with the analysis of Section 2, the total gravitational potential energy due to the ring and point mass will be determined first to find the gravitational force. The gravitational potential energy $\varphi$ between the point mass $M$ and the ring $\Omega$ can be then defined in general form as

$$\varphi = -GM \int_{\Omega} \frac{dm}{|r' + \xi|}, \quad (A1)$$

where $G$ is the universal gravitation constant, $dm$ is an infinitesimal mass element within the ring domain $\Omega$ and $r' + \xi$ is the vector connecting the mass element within the ring to the point mass $M$, as shown in Fig. A1. Since the ring normal is fixed and directed along the $e_3$ axis, the vector connecting the ring centre to the mass element can be defined as $\xi = (R\cos\phi, R\sin\phi, 0)$. Noting that $r' = r - \bar{r}$, equation (A1) can be written as

$$\varphi = -GM \int_{\Omega} \frac{dm}{\sqrt{(r-\bar{r})^2 + \xi^2 + 2\xi \cdot (r-\bar{r})}}, \quad (A2)$$

where $\xi^2 = R^2$ and $\xi \cdot (r-\bar{r}) = R\cos\phi(r-\bar{r}) \cdot e_1 + R\sin\phi(r-\bar{r}) \cdot e_1$. Furthermore, noting that $dm = \lambda R d\phi$, for a ring of uniform linear density $\lambda$, it can be shown that the gravitational potential energy $\varphi$ can be written as

$$\varphi(\bar{r}, r) = -\frac{GM\lambda}{k} \int_0^{2\pi} \frac{d\phi}{\sqrt{1 + k_1\cos\phi + k_2\sin\phi}}, \quad (A3)$$

where the non-dimensional parameters $k = \sqrt{1 + (r-\bar{r})^2/R^2}$, $k_1 = 2(r-\bar{r}) \cdot e_1/Rk^2$, and $k_2 = 2(r-\bar{r}) \cdot e_2/Rk^2$ are found from equation (A2) and are a function of the location of the centre of the ring $C$ and the location of the point mass $M$. It can be seen that as $r \to \bar{r}$, when the ring centre $C$ is coincident with the point mass $M$, that $k \to 1$ and $k_1, k_2 \to 0$ and so from equation (A3) the gravitational potential energy $\varphi \to -GMm/R$, as expected from Schumayer and Hutchinson (2019). For $\mathbf{r} \neq \bar{\mathbf{r}}$ integrating equation (A3) it can then be shown that the gravitational potential energy due to the point mass $M$ at position $\bar{r}$ and ring $\Omega$ with centre at position $r$ is given by

$$\varphi(\bar{\mathbf{r}}, \mathbf{r}) = -\frac{2GMm}{\pi R} \frac{1}{k\sqrt{1-\gamma}} K(\Lambda), \quad \gamma = \sqrt{k_1^2 + k_2^2}, \tag{A4}$$

where $K(\Lambda)$ is a complete elliptical integral of the first kind with argument $\Lambda = 2\gamma/(\gamma - 1)$, and the mass of the ring $m$ is given by $m = 2\pi \lambda R$, where $m \ll M$. Compared to Lass and Blitzer (1983), this more general expression ensures that the potential energy can be determined for an arbitrary ring location $r$ relative to a point mass at some position $\bar{r}$. The gravitational force $f(\bar{\mathbf{r}}, \mathbf{r})$ experienced by the ring due to the mass $M$ can finally be written as

$$f(\bar{\mathbf{r}}, \mathbf{r}) = -\nabla \varphi(\bar{\mathbf{r}}, \mathbf{r}) = \frac{2GMm}{\pi R} \frac{\partial}{\partial \mathbf{r}} \left[ \frac{1}{k\sqrt{1-\gamma}} K(\Lambda) \right], \tag{A5}$$

where $\nabla(.) = \partial(.)/\partial \mathbf{r}$. In order to proceed, the gravitational potential of the ring can now be defined as $\Phi(\bar{\mathbf{r}}, \mathbf{r}) = 1/m\varphi(\bar{\mathbf{r}}, \mathbf{r})$ to allow calculation of the gravitational acceleration terms in Section 2.

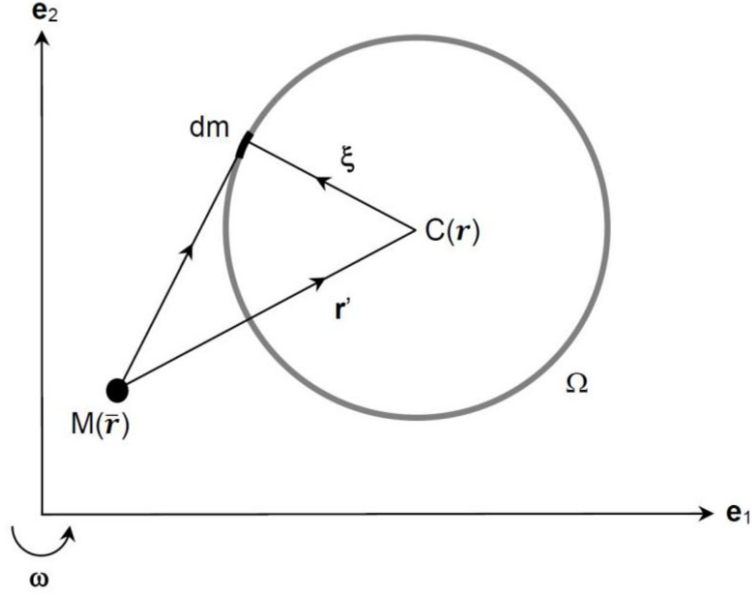

**Figure A1.** Uniform ring of radius $R$ with centre $C$ located at position $r$ and point mass $M$ located at position $\bar{r}$.

The gradient of the potential can also be expanded using the chain rule so that:

$$\nabla \varphi(\bar{\mathbf{r}}, \mathbf{r}) = \frac{\partial \varphi}{\partial \Lambda} \frac{\partial \Lambda}{\partial \mathbf{r}} + \frac{\partial \varphi}{\partial \gamma} \frac{\partial \gamma}{\partial \mathbf{r}} + \frac{\partial \varphi}{\partial k} \frac{\partial k}{\partial \mathbf{r}}, \tag{A6}$$

where the first partial derivative terms are given by

$$\frac{\partial \varphi}{\partial \Lambda} = \frac{GMm}{\pi R} \frac{(E(\Lambda) - K(\Lambda)(1-\Lambda))}{k\sqrt{1-\gamma}(1-\Lambda)\Lambda} \tag{A7a}$$

$$\frac{\partial \varphi}{\partial \gamma} = \frac{GMm}{\pi R} \frac{K(\Lambda)}{k(1-\gamma)^{3/2}}, \tag{A7b}$$

$$\frac{\partial \varphi}{\partial k} = \frac{2GMm}{\pi R} \frac{K(\Lambda)}{k^2\sqrt{1-\gamma}} \tag{A7c}$$

and where $E(\Lambda)$ is a complete elliptic integral of the second kind with argument $\Lambda$. The peculiarities and symmetries of the gravitational potential of a ring are discussed in detail by Schumayer and Hutchinson (2019).

**Centripetal and Coriolis forces**

In addition to the gravitational potential energy, and the resulting gravitational force experienced by the ring, the centripetal and Coriolis forces experienced by the ring in the rotating frame of reference can be determined. In general, consider now the ring, or any an extended rigid body, as comprising $N$ mass elements $m_i$ at position $\mathbf{r}_i (i = 1 - N)$. The location of the centre of mass is therefore defined as $\mathbf{r} = \sum m_i \mathbf{r}_i / \sum m_i$. First, the total centripetal force acting on the ring is given by $\sum m_i (\boldsymbol{\omega} \times \boldsymbol{\omega} \times \mathbf{r}_i)$, which can be written as $\boldsymbol{\omega} \times \boldsymbol{\omega} \times \sum m_i \mathbf{r}_i$ and so the total centripetal force is given by $m(\boldsymbol{\omega} \times \boldsymbol{\omega} \times \mathbf{r})$, where $m = \sum m_i$ is the total mass of the ring. Similarly, the Coriolis force is given by $2 \sum m_i (\boldsymbol{\omega} \times \dot{\mathbf{r}}_i)$, which can also be written as $2\boldsymbol{\omega} \times \sum m_i \dot{\mathbf{r}}_i$ and so the total Coriolis force is given by $2m(\boldsymbol{\omega} \times \dot{\mathbf{r}})$. The centripetal and Coriolis forces experienced by the ring are therefore equivalent to that experienced by a mass $m$ located at the centre of the ring, as utilized in Section 2.

## RING-RESTRICTED TWO-BODY PROBLEM

In order to provide insight into the stability analysis of the ring-restricted three-body problem presented in Section 5, the ring-restricted two-body problem can be considered as the limiting case when the mass ratio of the problem $\mu \to 0$. The ring-restricted two-body problem can then be defined using plane polar coordinates such that:

$$\frac{d^2 r(t)}{dt^2} - r(t)\omega(t)^2 = -\frac{\partial \Phi(r; R)}{\partial r} \tag{B1a}$$

$$\frac{1}{r(t)} \frac{d}{dt} \left[ r(t)^2 \omega(t) \right] = 0, \tag{B1b}$$

where $r(t)$ is the orbit radius defined by the location of the centre of the ring $C$, $\omega(t)$ is the orbital angular velocity of the ring and $\Phi(r; R)$ is the gravitational potential for a ring of non-dimensional ring radius $R$, where $R < r$. The centripetal term $r(t)\omega(t)^2$ is appropriate since it was demonstrated in Appendix A that the centripetal acceleration experienced by the ring is equivalent to that of a point mass located at the centre of the ring. From equation (B1b) it can be seen that the angular momentum per unit mass given by $h = r(t)^2 \omega(t)$ is conserved so that equation (B1a) can be reduced to:

$$\frac{d^2 r(t)}{dt^2} - \frac{h^2}{r(t)^3} = -\frac{\partial \Phi(r; R)}{\partial r}. \quad (B2)$$

Assuming now that $R \neq 0$, equation (B2) provides a general condition for the existence of a circular orbit of radius $r$ for a ring of radius $R$. For a circular orbit such that $r(t) = \tilde{r}$, it is now required that:

$$h^2 - \tilde{r}^3 \frac{\partial \Phi(r; R)}{\partial r}\bigg|_{r=\tilde{r}} = 0 \quad (B3)$$

In order to provide insights into the stability analysis presented in Section 5 in the limit that $\mu \to 0$, the conditions for stable and unstable two-body circular orbits for a ring will be investigated. First, in order to allow comparison the analysis of Section 5 in the rotating frame of reference, the orbital angular velocity will be fixed such that $\omega(t) = 1$. In general $h = r(t)^2 \omega(t)$, however the condition $\omega(t) = 1$ has now been defined and $r(t) = \tilde{r}$, so that $h = \tilde{r}^2$. Then, for a given ring radius $R$, equation (B3) can be solved to obtain the circular orbit radius $\tilde{r}$. As the ring radius increases, the required circular orbit radius also increases to ensure that the constraint $\omega(t) = 1$ is enforced, as can be seen in Fig. B1. For example, while $\tilde{r} \to 1$ as $R \to 0$ for a point mass two-body orbit, it is found from equation (B3) that when $R = 1$ the required circular orbit radius is now $\tilde{r} = 1.28$. It can also be noted from Fig. B1 that a collision with the central mass does not occur since $r > R$.

Now that the condition for a circular orbit for the ring has been determined, the stability properties of the orbit can be investigated. From equation (B2), an infinitesimal perturbation $\varepsilon(t)$ can be added to the circular orbit of radius $\tilde{r}$ such that $r(t) \to \tilde{r} + \varepsilon(t)$. Then, linearizing equation (B2) it can be shown that:

$$\frac{d^2 \varepsilon(t)}{dt^2} - \frac{\partial}{\partial r}\left[\frac{h^2}{r^3} - \frac{\partial \Phi(r; R)}{\partial r}\right]\bigg|_{r=\tilde{r}} \varepsilon(t) = 0 \quad (B4)$$

where the linear coefficient can be defined as $\Lambda^2 = \partial/\partial r [h^2/r^3 - \partial \Phi(r; R)/\partial r]|_{r=\tilde{r}}$. Again, since the condition $\omega(t) = 1$ has been selected, $h = \tilde{r}^2$ for an arbitrary circular orbit. Moreover, for linearly stable circular orbits; it is required that $\Lambda^2 < 0$, while unstable orbits are characterized by $\Lambda^2 > 0$. The form of $\Lambda$ can be determined and is shown in Fig. B1, where a transition from stable to unstable motion occurs at a ring radius $R_M = 0.705$ with the creation of a pair of real eigenvalues. This corresponds to the limiting case as $\mu \to 0$ in the full ring-restricted three-body problem, as discussed in Section 5.

Moreover, it can be noted that for a static ring encompassing the central mass then $h = 0$. From equation (B4), the condition for equilibrium is now $\partial \Phi(r; R)/\partial r|_{r \to 0} = 0$, which is satisfied if $r = 0$. This corresponds to a ring in a static equilibrium with the ring centre superimposed on the central mass. The stability properties of this static equilibrium configuration are then found from equation (B5) using:

$$\Lambda^2 = \lim_{r \to 0} \frac{\partial}{\partial r}\left[-\frac{\partial \Phi(r; R)}{\partial r}\right] = \frac{1}{2R^3}. \quad (B5)$$

Therefore, since $R > 0$ it is clear that $\Lambda^2 > 0$ and so the static equilibrium solution is always unstable, as expected from Maxwell (1859). The specific instability timescale for the ring dynamics is defined by $\Lambda^{-1}$ is also found in the analysis of McInnes (2003).

Finally, as the ring radius becomes small, and so the ring is tightly enclosing the smaller of the primary masses, it can be seen from Fig. 14 that the real eigenvalues grow rapidly, as expected from equation (B5). In this case the ring and the smaller of the primary masses form an approximate two-body pair. Then, comparing Fig. 14 and Fig. B1, it can be seen that as the ring radius becomes large the transition to instability occurs with the creation of pairs of real eigenvalues in both cases. Since the mass ratio of the problem is small for the Sun–Saturn case in Fig. 14, the transitions occur at a similar ring radius close to $R_M$, as seen in Fig. B1. In this case the ring and the larger of the primary masses form an approximate two-body pair.

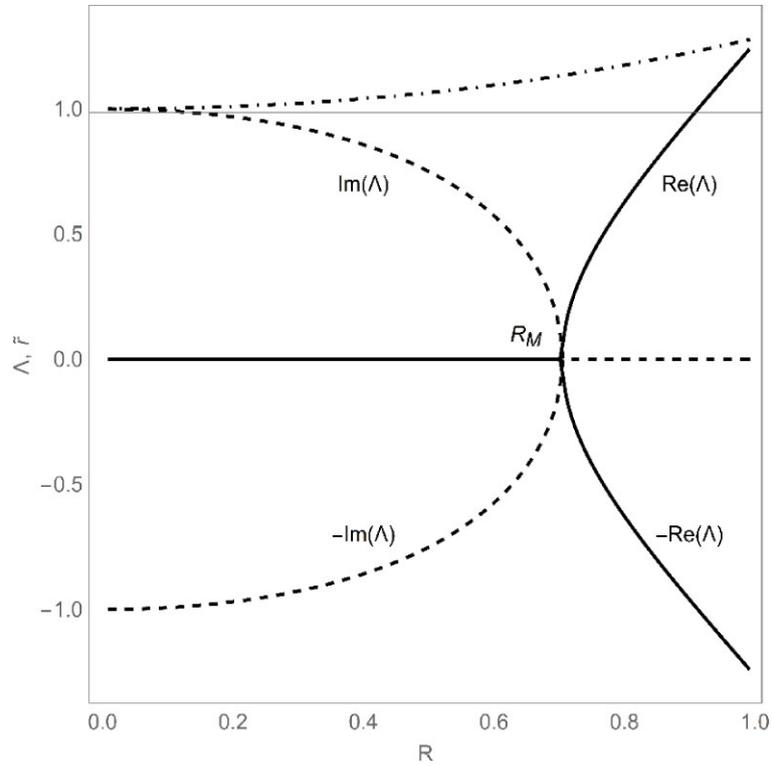

**Figure B1.** Eigenvalues Λ for the ring-restricted two-body problem. A linearly stable circular orbit of radius $\tilde{r}$ (- ·—·— · -) is obtained when $R < R_M$, after which a pair of real eigenvalues are created (—— real part, — — - imaginary part).

This paper has been typeset from a Microsoft Word file prepared by the author.